\documentclass[final,5p,times,twocolumn]{elsarticle}

\usepackage{amsmath}
\usepackage{amssymb}
\usepackage{color}
\usepackage{graphicx}
\usepackage{hyperref}
\usepackage[mathscr,scaled=1.15]{urwchancal}
\DeclareFontFamily{OT1}{pzc}{}
\DeclareFontShape{OT1}{pzc}{m}{it}%
{<-> s * [1.15] pzcmi7t}{}
\DeclareMathAlphabet{\mathpzc}{OT1}{pzc}{m}{it}

\definecolor{purple}{rgb}{0.5,0,0.5}
\definecolor{blue}{rgb}{0.0,0,0.9}

\biboptions{sort&compress}

\journal{Physics Letters B}

\begin{document}

\begin{frontmatter}

\title{Ward-Green-Takahashi identities and the axial-vector vertex}

\author[UFM]{Si-Xue Qin}
\author[ANL]{Craig D. Roberts}
\author[JARA]{Sebastian M. Schmidt}

\address[UFM]{Institut f\"{u}r Theoretische Physik, Johann
Wolfgang Goethe University, Max-von-Laue-Str.\,1, D-60438 Frankfurt am Main, Germany}
\address[ANL]{Physics Division, Argonne National Laboratory, Argonne, Illinois 60439, USA}
\address[JARA]{Institute for Advanced Simulation, Forschungszentrum J\"ulich and JARA, D-52425 J\"ulich, Germany}

\date{28 January 2014}

\begin{abstract}
The colour-singlet axial-vector vertex plays a pivotal role in understanding dynamical chiral symmetry breaking and numerous hadronic weak interactions, yet scant model-independent information is available.  We therefore use longitudinal and transverse Ward-Green-Takahashi (WGT) identities, together with kinematic constraints, in order to ameliorate this situation and expose novel features of the axial vertex: amongst them, Ward-like identities for elements in the transverse piece of the vertex, which complement and shed new light on identities determined previously for components in its longitudinal part.  Such algebraic results are verified via solutions of the Bethe-Salpeter equation for the axial vertex obtained using two materially different kernels for the relevant Dyson-Schwinger equations.  The solutions also provide insights that suggest a practical \emph{Ansatz} for the axial-vector vertex.
\end{abstract}


\end{frontmatter}

\noindent\textbf{1.$\;$Introduction}.\\
Dynamical chiral symmetry breaking (DCSB) is a crucial emergent phenomenon in the Standard Model, which can be characterised as the generation of \emph{mass from nothing} \cite{national2012Nuclear}.  It is a theoretically well-established feature of nonperturbative quantum chromodynamics (QCD) that, \emph{inter alia}, provides an explanation for both the proton's O$(1)$\,GeV mass in terms of strong-interaction dressing of the O$(1)\,$MeV current-quark masses and, simultaneously, how the pion nevertheless remains almost massless, on the hadronic scale, despite that hundred-fold magnification of the current masses \cite{Chang:2011vu,Bashir:2012fs,Cloet:2013jya}.

Drawing upon the foundation established by current-algebra and the hypothesis of partial conservation of the axial-current (PCAC) \cite{CAandA}, the keystone for understanding DCSB may be identified as the colour-singlet axial-vector vertex,\footnote{We focus herein on those components of $\Gamma_{5\mu}$ that are free from anomalies; i.e., what may be called the flavour-nonsinglet pieces.  Identities involving the  anomalous component are considered elsewhere \protect\cite{He:2002jg}.}  $\Gamma_{5\mu}$, which is the solution of a Bethe-Salpeter equation (BSE) with a $\gamma_5\gamma_\mu$ inhomogeneity.  In QCD, this vertex describes the nature of all measurable correlations between a dressed-quark and anti-quark that have nonzero overlap with the $J^{P}=1^{+}$ channel.  It is integral to comprehending hadronic weak interactions \cite{Hellstern:1997pg,Chang:2008sp,Chang:2009at}, in the same sense as its parity partner, $\Gamma_{\mu}$, the Schwinger function describing correlations in the $J^{P}=1^{-}$ channel, is crucial to explaining the electromagnetic interactions of hadrons \cite{Roberts:1994hh}.

Regarding $\Gamma_{\mu}$, a great deal of model-independent information has been garnered over a thirty-year period, as may be seen, e.g, from Refs.\,\cite{Ball:1980ay,Curtis:1990zs,Hawes:1991qr,
Burden:1993gy,Roberts:1994dr,Dong:1994jr,Bashir:1994az,Kizilersu:1995iz,Bashir:1995qr,
Kondo:1996xn,Bashir:1997qt,Walker:1999bp,He:1999hb,Davydychev:2000rt,
Skullerud:2003qu,Skullerud:2004gp,Boucaud:2003dx,Bashir:2004hh,Bhagwat:2004kj,
Bashir:2004mu,Pennington:2005mw,He:2006ce,Kizilersu:2009kg,He:2009sj,Binosi:2009qm,
Chang:2010hb,Bashir:2011dp,Qin:2013mta,Kizilersu:2013hea}.  It was accumulated using tools that range from perturbation theory to lattice gauge theory, and constraints such as the Ward-Green-Takahashi (WGT) identities \cite{Ward:1950xp,Green:1953te,Takahashi:1957xn,Takahashi:1985yz} and the Landau-Khalatnikov-Fradkin transformations \cite{LK56,Fradkin:1955jr,Johnson:1959zz,Zumino:1959wt}.  Notably, following the pattern set by Ref.\,\cite{Roberts:1994hh}, this knowledge has been invaluable in developing the theory and phenomenology of the spectrum and electromagnetic interactions of hadrons \cite{Chang:2011vu,Bashir:2012fs,Cloet:2013jya}.

\begin{figure}[t]
\centerline{%
\includegraphics[clip,width=0.45\linewidth]{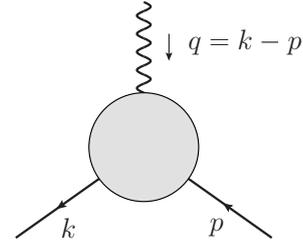}}
\caption{\label{figvertex} Axial-vector vertex, $\Gamma_{5\mu}(k,p)$, with the momentum flow indicated.  Plainly, $q=k-p$, and we define $t=(k+p)/2$ for later use.}
\end{figure}

On the other hand, available information on $\Gamma_{5\mu}$ is far less robust.  In fact, the difficulties with implementing chiral symmetry in lattice-QCD hamper  progress in that direction \cite{Boucaud:2009kv}, the \emph{Ans\"atze} used in hadron phenomenology are typically no more sophisticated than that written in Ref.\,\cite{Delbourgo:1979me}, the most complete computations \cite{Zong:2001im,Eichmann:2011pv,Chang:2012cc} are performed in the leading-order truncation of the Dyson-Schwinger equations (DSEs), and the impact of DCSB-generated nonperturbative corrections to the DSE kernels is only beginning to be explored and explicated \cite{Chang:2011ei}.  We describe remedial progress herein.

\medskip

\noindent\textbf{2.$\;$WGT Identities}.\\
We begin with the Ward-Green-Takahashi identities \cite{Ward:1950xp,Green:1953te,Takahashi:1957xn,Takahashi:1985yz} and focus on systems involving valence-quarks with degenerate current-masses: the generalisation to unequal current masses is straightforward.  The longitudinal identity is well known\footnote{We use a Euclidean metric:  $\{\gamma_\mu,\gamma_\nu\} = 2\delta_{\mu\nu}$; $\gamma_\mu^\dagger = \gamma_\mu$; $\gamma_5= \gamma_4\gamma_1\gamma_2\gamma_3$, tr$[\gamma_5\gamma_\mu\gamma_\nu\gamma_\rho\gamma_\sigma]=-4 \epsilon_{\mu\nu\rho\sigma}$; $\sigma_{\mu\nu}=(i/2)[\gamma_\mu,\gamma_\nu]$; $a \cdot b = \sum_{i=1}^4 a_i b_i$; and $q_\mu$ spacelike $\Rightarrow$ $q^2>0$.}
\begin{eqnarray}
q_\mu\Gamma_{5\mu}(k,p)+2im\Gamma_5(k,p)= S^{-1}(k)i\gamma_5+i\gamma_5 S^{-1}(p)\,,
\label{eq:longsA}
\end{eqnarray}
where $m$ is the current-quark mass,
\begin{equation}
S(k) = 1/[i\gamma\cdot k A(k^2) + B(k^2)]  \label{eqSk}
\end{equation}
is the dressed-quark propagator and $\Gamma_5(k,p)$ is the pseudoscalar vertex, obtained from a BSE defined with a $\gamma_5$ inhomogeneity.

The transverse identities \cite{Takahashi:1985yz,Kondo:1996xn,He:2000we,He:2006my,He:2007zza} are less familiar and, whereas the longitudinal WGT identity expresses properties of the divergence of the vertex, the transverse identities relate to its \emph{curl} (as Faraday's law of induction involves an electric field).  The transverse identities of use herein are:
\begin{subequations}
\label{eqTWGTIs}
\begin{eqnarray}
\nonumber \lefteqn{q_\nu \Gamma_{\mu}(k,p) - q_\mu \Gamma_{\nu}(k,p) =
S^{-1}(k)\sigma_{\mu\nu} + \sigma_{\mu\nu} S^{-1}(p)}\\
&& - 2m\Gamma_{\mu\nu}(k,p) - 2\,t_\lambda\varepsilon_{\lambda\mu\nu\rho}\Gamma_{5\rho}(k,p)
+ A^V_{\mu\nu}(k,p)\,, \label{eqTWGTI}\\
\nonumber
\lefteqn{
q_\mu \Gamma_{\nu\rho}(k,p) + q_\nu \Gamma_{\rho\mu}(k,p) + q_\rho \Gamma_{\mu\nu}(k,p) = S^{-1}(k)\varepsilon^5_{\mu\nu\rho} }\\
&& - \varepsilon^5_{\mu\nu\rho}S^{-1}(p)
+ 2it_\lambda\varepsilon_{\lambda\mu\nu\rho}\Gamma_5(k,p)
 + P_{\mu\nu\rho}(k,p)\,, \label{eqTAWGTIP}
\end{eqnarray}
\end{subequations}
where $\varepsilon^5_{\mu\nu\rho}=\varepsilon_{\mu\nu\rho\lambda}\gamma_\lambda
\gamma_5$ and $\Gamma_{\mu\nu}(k,p)$ is a rank-$2$ tensor vertex, obtained from a BSE defined with a $\sigma_{\mu\nu}$ inhomogeneity.

The last two terms in Eq.\,\eqref{eqTWGTI} arise in computing the momentum space expression of a nonlocal axial-vector vertex, whose definition involves a gauge-field-dependent line integral \cite{He:2007zza}; and the last two terms in Eq.\,\eqref{eqTAWGTIP} arise from similar manipulations of an analogous nonlocal tensor vertex.  Note that, like Eq.\,\eqref{eq:longsA}, the transverse identities are valid in any covariant gauge, which is the class we focus upon, and do not explicitly display dependence on the gauge-fixing parameter.  It is straightforward to verify that Eqs.\,\eqref{eq:longsA}\,--\,\eqref{eqTWGTIs} are satisfied by the bare propagators and vertices, given that $A^V_{\mu\nu}(k,p)$, $P_{\mu\nu\rho}(k,p)$ are zero in the absence of interactions.

The presence of the unfamiliar vertices $A^V_{\mu\nu}(k,p)$, $P_{\mu\nu\rho}(k,p)$ in Eqs.\,\eqref{eqTWGTIs} lends them an appearance of impracticality, since even at one-loop order the expressions for such quantities are complicated \cite{He:2002jg,Pennington:2005mw,He:2006ce} and, moreover, they lead to a coupling between the WGT identities.  Notwithstanding these ostensible difficulties, progress can be made in the absence of detailed forms for $A^V_{\mu\nu}(k,p)$, $P_{\mu\nu\rho}(k,p)$ following the approach of Ref.\,\cite{Qin:2013mta}. One first defines tensors
\begin{eqnarray}
T^1_{\mu\nu}=\mbox{\footnotesize $\displaystyle \frac{1}{4}$}
\varepsilon_{\alpha\mu\nu\beta}t_\alpha q_\beta\, \mathbf{I}_{\rm D},
\quad
T^2_{\mu\nu}= \mbox{\footnotesize $\displaystyle \frac{1}{4}$}
\varepsilon_{\alpha\mu\nu\beta}\gamma_\alpha q_\beta,
\end{eqnarray}
where $\mathbf{I}_{\rm D}$ is the $4\times 4$ identity matrix in spinor space.  Then, following their separate contraction with Eq.\,\eqref{eqTWGTI}, the contraction of Eq.\,\eqref{eqTAWGTIP} with $q_\rho$, and elimination of $q^2 T_{\mu\nu}^{1,2} \Gamma_{\mu\nu}(k,p)$ from the equations obtained in favour of terms with no explicit involvement of $\Gamma_{\mu\nu}(k,p)$, one arrives at the following identities:
\begin{eqnarray}
q^2\left[ t^2\,q_\mu \Gamma_{5\mu}(k,p) - q\cdot t\,t_\mu\Gamma_{5\mu}(k,p)\right]
&=& \mathpzc{R}^1(k,p)\,, \label{eq:transA31}\\
q^2\left[ \gamma\cdot t \,q_\mu\Gamma_{5\mu}(k,p) - q\cdot t\,\gamma_\mu\Gamma_{5\mu}(k,p)  \right] &=& \mathpzc{R}^2(k,p)\,,
\label{eq:transA32}
\end{eqnarray}
with
{\allowdisplaybreaks
\begin{eqnarray}
\nonumber
\lefteqn{
\mathpzc{R}^{1,2}(k,p) =
q^{2} T^{1,2}_{\mu\nu}\left[S^{-1}(k)\sigma_{\mu\nu} + \sigma_{\mu\nu} S^{-1}(p)\right]}  \\
\nonumber
&& -  \, 2m  T^{1,2}_{\mu\nu}\left[S^{-1}(k)\varepsilon^5_{\mu\nu\rho}q_\rho
- \varepsilon^5_{\mu\nu\rho} q_\rho S^{-1}(p)  \right. \\
&&  \left. \rule{0em}{2ex}
+ 2it_\lambda q_\rho \varepsilon_{\lambda\mu\nu\rho}\Gamma_5(k,p)\right]
+ T^{1,2}_{\mu\nu}Y_{\mu\nu}(k,p) ,
\end{eqnarray}}
\hspace*{-0.5\parindent}where $Y_{\mu\nu}(k,p) = q^2 A^V_{\mu\nu}(k,p)- {2m}\,q_\rho{P}_{\mu\nu\rho}(k,p)$ denotes all higher-order terms.  Notably, the axial-vector and pseudoscalar vertices remain coupled, even in the chiral limit.  If this were not the case, then current algebra and the PCAC hypothesis could not have been successful.

With Eqs.\,\eqref{eq:longsA}, \eqref{eq:transA31}, \eqref{eq:transA32} one has a set of three matrix-valued identities for scalar-valued projections of $\Gamma_{5\mu}(k,p)$, which amounts to twelve linearly-independent but coupled linear equations for twelve unknown scalar functions.  One complication remains; namely, the contractions $T^{1,2}_{\mu\nu} Y_{\mu\nu}(k,p)$.  They are computable once a truncation for the associated DSEs is decided upon.  However, a detailed form is not immediately necessary.  One can proceed by noting that they are merely matrix-valued scalar amplitudes and hence can be expressed succinctly:
\begin{eqnarray}
iT^{1,2}_{\mu\nu}Y_{\mu\nu}(k,p)&=&\sum_{i=1}^{4} \mathcal{P}_i(k,p) Y_{i}^{1,2}(k,p),
\end{eqnarray}
where $\{\mathcal{P}_i, i=1,\ldots,4\}$ is the pseudoscalar Dirac-matrix basis
\begin{equation}
\begin{array}{ll}
\mathcal{P}_1(k,p)= \gamma_5\,, & \mathcal{P}_2(k,p)=\gamma_5(\gamma\cdot q)\,,\\
\mathcal{P}_3(k,p)=\gamma_5(\gamma\cdot t)(q\cdot t), &
\mathcal{P}_4(k,p)=\gamma_5\, \sigma_{\alpha\beta}\,
 t_\alpha \, q_\beta\,,
\end{array}
\end{equation}
and the elements $\{Y_{i}^{1,2}, i=1,\ldots,4\}$ are scalar functions, to be determined.  At this point, a solution of Eqs.\,\eqref{eq:longsA}, \eqref{eq:transA31}, \eqref{eq:transA32} provides an expression for $\Gamma_{5\mu}(k,p)$ in terms of the dressed-quark propagator, the pseudoscalar vertex and $\{Y_{i}^{1,2}, i=1,\ldots,4\}$.
%
%
%
%
%
%
%
%

\medskip

\noindent\textbf{3.$\;$Algebraic solution of the coupled identities}.\\
Retaining complete generality, one may write
\begin{eqnarray}
\nonumber
\lefteqn{i\Gamma_{5}(k,p)= i\mathcal{P}_1(k,p) E_5 (k,p) + \mathcal{P}_2(k,p) F_5 (k,p)}\\
& &\quad \quad + \mathcal{P}_3(k,p) G_5(k,p) +\mathcal{P}_4(k,p) H_5(k,p)\,,\\
\lefteqn{\!\!\Gamma_{5\mu}(k,p) = \Gamma_{5\mu}^\lambda(k,p)+ \Gamma_{5\mu}^\tau(k,p)}\\
&=&\!\!\!\! \sum_{j=1}^4 \lambda_j(k,p)\, \gamma_5\, L^j_\mu(k,p)  + \sum_{j=1}^8 \tau_j(k,p) \, \gamma_5\, T^j_\mu(k,p)\,,\quad \label{eqDecomposeAV}
\end{eqnarray}
where the longitudinal matrix-valued tensors are $(\hat q\cdot\hat q=1)$
\begin{equation}
\label{eqLBasis}
\begin{array}{ll}
L^1_\mu(k,p) = i q_\mu \mathbf{I}_{\rm D}\,, &
L^2_\mu(k,p) = \hat q_\mu (\gamma\cdot \hat q) \,\\
L^3_\mu(k,p) = \hat q_\mu (\gamma\cdot t) \,,&
L^4_\mu(k,p) = \hat q_\mu (\sigma_{\alpha\beta}\, t_\alpha \,\hat q_\beta) \,,
\end{array}
\end{equation}
and the tensors used to express the transverse part of the axial-vector vertex, $\{T^j_\mu(k,p),j=1,\ldots,8\}$, are defined in Eq.\,\eqref{eqTransverse}.  One can obtain the solution of Eqs.\,\eqref{eq:longsA}, \eqref{eq:transA31}, \eqref{eq:transA32} using any reliable means to solve a system of coupled linear equations.

A first noteworthy result is that, irrespective of the nature of $Y_{\mu\nu}$, the longitudinal piece of the axial vertex, $\Gamma_{5\mu}^\lambda$, is completely determined by the dressed-quark propagator and the pseudoscalar vertex; viz., with $\Sigma_\phi(k^2,p^2):=[\phi(k^2)+\phi(p^2)]/2$, $\Delta_{\phi}(k^2,p^2):=[\phi(k^2)-\phi(p^2)]/[k^2-p^2]$,
\begin{subequations}
\label{eqlambdas}
\begin{eqnarray}
\lambda_1(k,p)&=&\frac{1}{q^2}\left[2 \Sigma_B(k^2,p^2)-2m E_5(k,p) \right]\,,\\
\lambda_2(k,p)&=&\Sigma_A(k^2,p^2)-2mF_5(k,p)\,,\\
\lambda_3(k,p)&=&2\,t\cdot \hat q  \left[\Delta_A(k^2,p^2)-mG_5(k,p) \right]\,,\\
\lambda_4(k,p)&=&-2m H_5(k,p)\,.
\end{eqnarray}
\end{subequations}

The solution for the transverse part, $\Gamma_{5\mu}^\tau$, is determined by the scalar functions $\{\tau_i,i=1,\ldots,8\}$.  In order to display the results, it is useful to write $\{\tau_i =: \tau^S_i + \tau^Y_i,i=1,\ldots\,8\}$, wherein the $\{\tau^S_i\}$ are fully determined by the scalar functions appearing in the dressed-quark propagator, Eq.\,\eqref{eqSk}; viz.,
\begin{subequations}
\label{eqtauiS}
\begin{eqnarray}
%
\tau_1^S(k,p)&=&\Sigma_A(k^2,p^2)+\mbox{\footnotesize $\displaystyle \frac{1}{2}$} q^2\Delta_A(k^2,p^2)\notag\\
&& + \, 2m\Delta_B(k^2,p^2), \label{eqtau1S}\\
\tau_3^S(k,p)&=& 2m\Delta_A(k^2,p^2)\,,\\
\tau_4^S(k,p)&=&\mbox{\footnotesize $\displaystyle \frac{1}{2}$} \Delta_A(k^2,p^2)\,,\\
\tau_5^S(k,p)&=& \frac{m\Sigma_A(k^2,p^2)-\Sigma_B(k^2,p^2)}{t\cdot q}\,, \label{eqtau5}\\
\tau_7^S(k,p)&=&2\Delta_A(k^2,p^2)\,,\\
\tau_2^S(k,p)&=& \tau_6^S(k,p) \;\; = \;\; \tau_8^S(k,p) \;\; = \;\; 0\,.
\end{eqnarray}
\end{subequations}


The scalar functions $\{\tau^Y_i\}$, on the other hand, express contributions from $\{Y_{i}^{1,2}, i=1,\ldots,4\}$, as apparent in Eqs.\,\eqref{eqtauY}.  Amongst this set, the function $\tau_5^Y$ presents a curious case.  Notwithstanding their strengths, the WGT-identities do not impose the physical constraint that $\Gamma_{5\mu}(k,p)$ should be free of kinematic singularities at $k^2=p^2$, which is the Ward-identity limit.  Plainly, $\tau_5^S(k,p)$ is singular at this point.  Therefore, $\tau_5^Y(k,p)$ must be nonzero in general and contain a piece, completely determined by the dressed-quark propagator, which exactly cancels the singularity.  We will return to this point.

\medskip

\noindent\textbf{4.$\;$Neighbourhood of a pseudoscalar meson pole}.\\[0.5ex]
\noindent\emph{4.1$\;$General observations}.
Consider the possibility that the longitudinal part of the axial-vector vertex possesses a simple pole, at $q^2+m_P^2=0$, associated with a pseudoscalar bound-state.  In that case, the pseudoscalar vertex must possess a similar pole; and in the neighbourhood $q^2+m_P^2\simeq 0$, one may write\footnote{In this section it is advantageous to use a basis for the non-transverse part of the vertex that differs from that in Eqs.\,\protect\eqref{eqLBasis}.  The mapping between Eqs.\,\protect\eqref{eqG5mu} and \protect\eqref{eqDecomposeAV} is straightforward.}
\begin{eqnarray}
\nonumber\lefteqn{
\Gamma_{5\mu}(k,p) = \sum_{j=1}^8 \overline{\tau}_j(k,p) \, \gamma_5\, T^j_\mu(k,p)}\\
\nonumber && + i\gamma_5\,t_\mu E_R (t,q) + \gamma_5 \gamma_\mu F_R (t,q)  + \gamma_5\,t_\mu(\gamma\cdot t)G_R(t,q)\\
&& -\gamma_5 \sigma_{\mu\nu}\,
 t_\nu H_R(t,q) + q_\mu \frac{2 r_A \Gamma_P(t,q)}{q^2+m_P^2}\,, \label{eqG5mu}\\
\nonumber
\lefteqn{\Gamma_{5}(k,p)= \mathcal{P}_1(k,p) E_5^R (t,q) - i\mathcal{P}_2(k,p) F_5^R (t,q)-i\mathcal{P}_3(k,p) }\\
&& \times \, G_5^R(t,q) - i\mathcal{P}_4(k,p) H_5^R(t,q) -\frac{2i r_P\Gamma_P(t,q)}{q^2+m_P^2}\, , \label{eqG5}\
\end{eqnarray}
where the putative pole is made explicit, so that each element remaining in Eqs.\,\eqref{eqG5mu}, \eqref{eqG5} is regular on $q^2+m_P^2\simeq 0$.  In these equations, $2 r_A$ and $2r_P$ are, respectively, the residues of the supposed pseudoscalar meson pole in the axial-vector and pseudoscalar vertices (the explicit factor of ``2'' reflects considerations associated with the flavour structure of the vertex and canonical normalisation of the bound-state \cite{Maris:1997hd,Maris:1997tm}), and
\begin{eqnarray}
\nonumber
\lefteqn{\Gamma_P(t,q)= i \mathcal{P}_1(k,p) E_P (t,q) + \mathcal{P}_2(k,p) F_P (t,q)}\\
& & + \mathcal{P}_3(k,p)G_P(t,q) + \mathcal{P}_4(k,p)  H_P(t,q)
\end{eqnarray}
is the pseudoscalar meson's Bethe-Salpeter amplitude \cite{LlewellynSmith:1969az}.

It is informative to substitute these decompositions into the WGT identities, Eqs.\,\eqref{eq:longsA}, \eqref{eq:transA31}, \eqref{eq:transA32}, an operation which yields the following relations, valid on $q^2+m_P^2\simeq 0$:
\begin{subequations}
\label{eqPointGB}
\begin{eqnarray}
\nonumber
E_R(t,q)\!\! &=& \!\!  \frac{2}{q\cdot t}\left[ \Sigma_B(k^2,p^2) - r_A E_P(t,q) \right.\\
 &&\left. -  mE_5^R(t,q) +  \mathcal{N}_P(q^2)E_P(t,q) \right],\\
F_R(t,q)\!\! &=& \!\!  \Sigma_A(k^2,p^2) - 2\,[ r_A F_P(t,q)\notag\\
 && + mF_5^R(t,q)- \mathcal{N}_P(q^2)F_P(t,q)]\,,\\
G_R(t,q)\!\! &=&\!\!  2\, [ \Delta_A(k^2,p^2)- r_{A}G_P(t,q)\notag\\
 && - mG_{5}^R(t,q)+ \mathcal{N}_P(q^2)G_P(t,q)]\,,\\
\nonumber
H_R(t,q)\!\! &=& \!\! 2\, [- r_A H_P(t,q)- m H_5^R(t,q)\\
&& + \mathcal{N}_P(q^2)H_P(t,q)]\,, \;\;
\end{eqnarray}
\end{subequations}
where
$\mathcal{N}_P(q^2) = [m_\pi^2 r_A - 2 m r_P]/[q^2+m_P^2]$.  Since the left-hand-side in each of these identities is regular by definition, then
\begin{equation}
\label{eqGMOR}
r_A m_P^2 = 2 m r_P\,.
\end{equation}
This is the generalised form of the Gell-Mann--Oakes--Renner relation derived in Ref.\,\cite{Maris:1997hd}, wherein it is also shown that $r_A$ is the pseudoscalar meson's leptonic decay constant, $r_A = f_P$, and the product $f_P r_P$ is the in-meson condensate \cite{Maris:1997tm,Brodsky:2010xf,Chang:2011mu,Brodsky:2012ku}.

Since Eq.\,\eqref{eqGMOR} is valid in the neighbourhood of any pseudoscalar meson pole, there are numerous corollaries and generalisations, some of which apply to heavy-light \cite{Ivanov:1998ms} and heavy-heavy \cite{Bhagwat:2006xi} pseudoscalar mesons, and others to radially excited and hybrid pseudoscalar mesons \cite{Holl:2004fr,Holl:2005vu,McNeile:2006qy}.  In the latter connection it is important to highlight one aspect of the corollary.  Consider the ground-state and suppose that chiral symmetry is dynamically broken, in which case $f_\pi \neq 0$.  Then, according to Eq.\,\eqref{eqGMOR}, it follows that the ground-state is massless in the chiral limit because $m r_P \equiv 0$.  On the other hand, in the same circumstances the mass of any non-ground-state pseudoscalar meson is nonzero; and hence Eq.\,\eqref{eqGMOR} entails $r_A=f_P=0$, so that the non-ground-state pseudoscalar meson pole disappears completely from the axial-vector vertex.
\smallskip

\noindent\emph{4.2$\;$Pseudoscalar meson ground state}.
Let us focus first, therefore, on the ground state in the chiral limit.  Then, on $q^2\simeq 0$, with $y:=t^2$, $w:= t\cdot q$, Eqs.\,\eqref{eqPointGB} entail the following Goldberger-Treiman-like identities
\begin{subequations}
\label{newGMOR}
{\allowdisplaybreaks
\begin{eqnarray}
E_R(y,w) & = & \frac{2}{w} \left[\Sigma_B(y+w,y-w) - f_P E_P(y,w)\right],\\
F_R(y,w)&=& \Sigma_A(y+w,y-w) - 2 f_P F_P(y,w)\,,\\
G_R(y,w) &=& 2\Delta_A(y+w,y-w) - 2 f_P G_P(y,w)\,,\\
H_R(y,w) & = & - 2 f_P H_P(y,w)\,.
\end{eqnarray}}
\end{subequations}

In the chiral limit, indeed, whenever the valence-quark constituents are degenerate, the pseudoscalar meson multiplet contains an eigenstate of the charge conjugation operator and hence $E_P(y,w)$ is an even function of $w$.  Using this feature in conjunction with the fact that, by construction, $E_R$ cannot possess a pole on $q^2\simeq 0$, then Eqs.\,\eqref{newGMOR} yield the relations obtained in Ref.\,\cite{Maris:1997hd} and verified in Ref.\,\cite{Maris:1997tm}; viz.,
\begin{subequations}
\label{OriginalGGTR}
\begin{eqnarray}
f_P E_P(y,w=0) &=& B(y)\,, \label{OriginalGTR}\\
F_R(y,w=0) + 2 f_P F_P(y,w=0) &=& A(y) \,,\\
G_R(y,w=0) + 2 f_P G_P(y,w=0) &=& 2A'(y)\,,\\
H_R(y,w=0) + 2 f_P  H_P(y,w=0)& =& 0\,,
\end{eqnarray}
\end{subequations}
and, plainly, $E_R(y,w=0)=0$.  These are Ward-like identities for the axial-vector vertex.

Equations\,\eqref{newGMOR} are more general than Eqs.\,\eqref{OriginalGGTR}, being valid at each value of $(y,w)$ on $q^2\simeq 0$.  Consequently, if one knows the dressed-quark propagator and ground-state pseudoscalar meson's bound-state amplitude in the chiral limit, then the longitudinal part of the axial-vector vertex is completely specified on this domain.

Notably, $B(y) \neq 0$ in the chiral limit is synonymous with DCSB. Equation\,\eqref{OriginalGTR} can thus be used to argue \cite{Maris:1997hd} that, in the chiral limit, DCSB is a sufficient and necessary condition for the appearance of a massless pseudoscalar bound state that is also the dominant feature of the axial-vector vertex on $q^2\simeq 0$.  (This is not true of the flavour-singlet component \cite{Bhagwat:2007ha}.)

\smallskip

\noindent\emph{4.3$\;$Pseudoscalar meson excited states}.
Equations.\,\eqref{eqPointGB}, with $\mathpzc{N}_P(q^2)\equiv 0$, are valid in the neighbourhood of the pole associated with any pseudoscalar meson constituted from valence-quarks with nonzero current-mass.  In the chiral limit, however, as noted in the paragraph preceding Sec.\,4.2, the pole associated with any non-ground-state pseudoscalar meson decouples from the axial-vector vertex, so steps used in the derivation of these equations are invalid.  Thus, in the chiral limit one may revert to Eqs.\,\eqref{eqlambdas}: inserting $q^2=-m_P^2$, they yield $(y_\pm = y \pm w -m_P^2/4)$
\begin{subequations}
\label{eqlambdaiE}
\begin{eqnarray}
\lambda_1(y,w;-m_P^2) &=& -\frac{2 }{m_P^2} \Sigma_B(y_+,y_-) \,,\\
\lambda_2(y,w;-m_P^2) &=& \Sigma_A(y_+,y_-) \,, \\
\lambda_3(y,w;-m_P^2) &=&  -\frac{2 w }{m_P^2} \Delta_A(y_+,y_-)\,,\\
\lambda_4(y,w;-m_P^2) & \equiv & 0\,.
\end{eqnarray}
\end{subequations}
Evidently, on $q^2+m_P^2\simeq 0$ the momentum-dependence of the longitudinal piece of the axial-vector vertex is completely determined by that of the dressed-quark propagator.

\medskip

\noindent\textbf{5.$\;$Transverse part of the axial-vector vertex}.\\
The presence or absence of a pseudoscalar meson pole has no effect on the transverse part of the axial-vector vertex, which is determined by the sum $\{\tau_i = \tau_i^S + \tau_i^Y,i=1,\ldots,8\}$ with $\{\tau_i^S,i=1,\ldots,8\}$ and $\{\tau_i^Y,i=1,\ldots,8\}$ given, respectively, in Eqs.\,\eqref{eqtauiS}, \eqref{eqtauY}.  It is worth understanding the role of the higher-order terms, $\{\tau_i^Y,i=1,\ldots,8\}$ so that, e.g., one may continue forming impressions that inform the construction of an \emph{Ansatz} for $\Gamma_{5\mu}$.  (Naturally, $\Gamma_{5\mu}^\tau(k,p)$ will exhibit resonance structures associated with each axial-vector meson but this feature is not germane to the present discussion.)

As we remarked in the last paragraph of Sec.\,3, $\Gamma_{5\mu}(t,q)$ must be free of kinematic singularities at $k^2-p^2=2 t\cdot q = 2 w = 0$, and hence one may write
\begin{equation}
\tau_5^Y(y,w;q^2) = -\tau_5^S(y,w;q^2) + \frac{w \tilde Y_1^1(y,w;q^2)}{w^2 - y\, q^2}\,,
\end{equation}
where $\tilde Y_1^1(y,w;q^2)$ is even under $w\to -w$ owing to the charge-conjugation symmetry of $\Gamma_{5\mu}(t,q)$.  Following this observation, we arrive at the following Ward-like identity
\begin{equation}
\label{eqtau5zero}
\tau_5(y,w=0;q^2)  \equiv 0\,.
\end{equation}


In order to elucidate these and related matters, we solved the vertex Bethe-Salpeter equation using the two distinct symmetry preserving kernels detailed in Appendix\,A of Ref.\,\cite{Chang:2012cc}, with a current-quark mass that produces $m_\pi = 0.14\,$GeV.  One solution is obtained using the rainbow-ladder (RL) truncation, which is the leading-order in a systematic, symmetry-preserving scheme \cite{Munczek:1994zz,Bender:1996bb}.  The other is obtained with the most sophisticated kernel that is currently available; namely, a DCSB-improved (DB) kernel that incorporates essentially nonperturbative effects generated by DCSB that are omitted in RL truncation and any stepwise improvement thereof \cite{Chang:2009zb,Chang:2010hb,Chang:2011ei}.

An immediate truncation-independent result is confirmation of Eq.\,\eqref{eqtau5zero}.  Moreover, the functions $\tau_{2,6,8}$, for which $\tau_{2,6,8}^S\equiv 0$, remain zero after the addition of $\tau_{2,6,8}^Y(y,w=0,q^2)$.  Hence, we obtain another set of Ward-like identities for elements in the transverse part of the axial-vector vertex:
\begin{equation}
\label{eqtau268zero}
\tau_{2,6,8}(y,w=0;q^2)  \equiv 0\,.
\end{equation}
This, again, is because kinematic singularities cannot appear; the tensors $T_\mu^{2,6,8}$ [Eqs.\,\eqref{eqT2}, \eqref{eqT6}, \eqref{eqT8}] are odd under the charge conjugation operation; and hence $\tau_{2,6,8}(y,w\simeq 0,q^2) \propto w$ in a charge-conjugation invariant vertex.

\begin{figure}[t]
\centerline{%
\includegraphics[clip,width=0.9\linewidth]{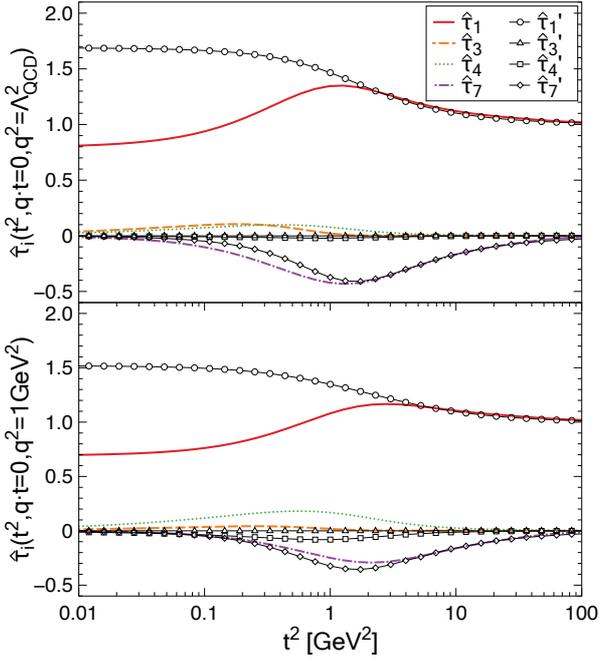}}
\caption{\label{FigRL} \emph{Rainbow-ladder truncation} (RL),  $\tau_{1,3,4,7}(y,w=0;q^2)$; viz., the only functions in the transverse piece of the axial-vector vertex that are nonzero at $w=0$.  N.B.\, The circumflex indicates that each function is multiplied by an appropriate power of $|t|$ in order to make it dimensionless; the primed quantities denote $\hat \tau_{1,3,4,7}^S(y,w=0;q^2)$; and $\Lambda_{\rm QCD} \approx 0.2\,$GeV.}
\end{figure}

\begin{figure}[t]
\centerline{%
\includegraphics[clip,width=0.9\linewidth]{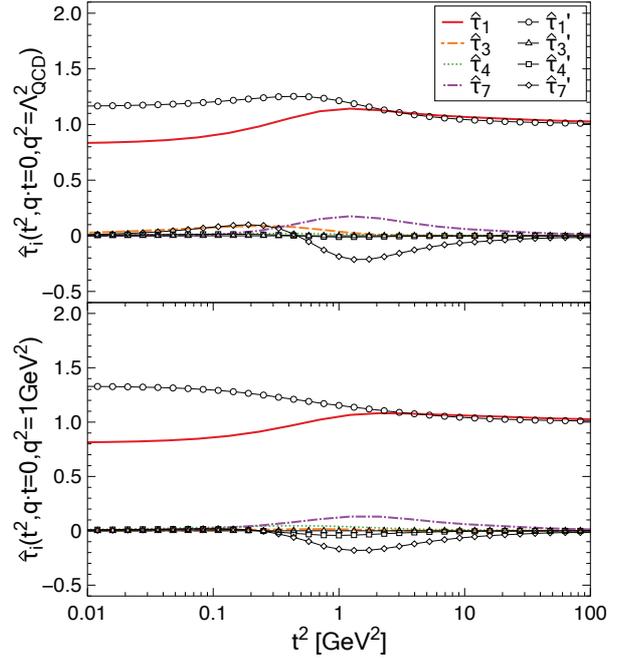}}
\caption{\label{FigDB} \emph{DCSB-improved truncation} (DB),
$\tau_{1,3,4,7}(y,w=0;q^2)$; viz., the only functions in the transverse piece of the axial-vector vertex that are nonzero at $w=0$.  N.B.\, The circumflex indicates that each function is multiplied by an appropriate power of $|t|$ in order to make it dimensionless; the primed quantities denote $\hat \tau_{1,3,4,7}^S(y,w=0;q^2)$; and $\Lambda_{\rm QCD} \approx 0.2\,$GeV.}
\end{figure}

We plot the functions $\tau_{1,3,4,7}(y,w=0;q^2)$ in Figs.\,\ref{FigRL} and \ref{FigDB}: comparing the panels reveals the effect of both improving the truncation and evolution with spacelike $q^2$.  Regarding the latter, each function's magnitude typically falls with increasing $q^2$.

As one would have anticipated, $\tau_1$, the coefficient of $\gamma_5\gamma_\mu^T$, is the dominant function in the transverse part of the axial-vector vertex,  independent of the DSE kernels; and it evolves at ultraviolet momenta, $t^2=y\gtrsim 1.5\,$GeV$^2$, according to Eq.\,\eqref{eqtau1S}.  Its behaviour at infrared momenta is sensitive to the truncation, with the DB-kernel producing a result that more closely tracks that in Eq.\,\eqref{eqtau1S}.  This is readily understood because it has long been known that in order to describe a given set of hadron physics observables with a RL-kernel, too much interaction strength must be located at infrared momenta, leading to magnifications of $A(p^2=0)$, $B(p^2=0)$ that are unrealistically large \cite{Maris:2002mt}.
With DB kernels, on the other hand, the effect of DCSB is expressed more realistically, being distributed over each of the elements that appear in the kernel's construction \cite{Chang:2009zb,Chang:2011ei}.  Consequently, the interaction need not be over-enhanced at infrared momenta, so that the nonperturbative dynamical contributions expressed in $\{Y_i^{1,2}, i=1,\ldots,4\}$ are smaller and the Schwinger functions evolve less rapidly with momenta in order to reach their ultraviolet limits.
%

The behaviour of the subleading functions, $\tau_{3,4,7}(y,w=0;q^2)$, fits the same pattern.  Each one is essentially nonperturbative, because it is associated with a tensor structure that does not appear in QCD's Lagrangian, and therefore vanishes as a power law at ultraviolet momenta.  The momentum-dependence of a given function at infrared momenta is sensitive to the structure of the kernel but the magnitude is smaller when the DB kernel is used.  The identities in Eqs.\,\eqref{eqtauiS} are a fair guide to the magnitudes of the functions $\tau_{3,4,7}(y,w=0;q^2)$.  However, they do not always predict the correct sign, which also depends on the kernel used.  This shows that corrections from $\tau_{3,4,7}^{Y}$ can be noticeable.  It follows that if one employs an \emph{Ansatz} based on Eqs.\,\eqref{eqtauiS}, then a reasonable error estimate may be obtained by exploring the response of a given result to changes in the sign of these terms.

%

\medskip

\noindent\textbf{6.$\;$Summary and Conclusions}.
Numerous symmetries of QCD's Lagrangian are expressed in the longitudinal and transverse Ward-Green-Takahashi (WGT) identities.  We used those identities, together with kinematic constraints, in order to expose novel features of the colour-singlet axial-vector vertex.

Simplest amongst our results are a set of Ward-like identities for elements in the transverse part of the axial-vector vertex, Eqs.\,\eqref{eqtau5zero}, \eqref{eqtau268zero}.  They complement identities determined previously for elements in the longitudinal part of the vertex, Eqs.\,\eqref{OriginalGGTR}, which may now be viewed from a new perspective.  We showed, too, that in the chiral limit, in the neighbourhood of the pole associated with any pseudoscalar meson excited-state, the momentum-dependence of the longitudinal part of the axial-vector vertex is completely determined by the dressed-quark propagator, Eqs.\,\eqref{eqlambdaiE}.

In addition, we solved the inhomogeneous Bethe-Salpeter equation for the axial-vector vertex using two materially different truncations of the relevant Dyson-Schwinger equation kernels.  This enabled us to verify all algebraic results.
The solutions also provided insights that suggest a form for the vertex which, in the neighbourhood of the chiral limit, may be employed usefully on $q^2\geq 0$ by those practitioners wishing to work simply with solutions of the gap equation, thereby overcoming the need for solving the inhomogeneous Bethe-Salpeter equation in addition; viz., $(y_\pm = t^2 \pm t\cdot q + q^2/4)$
{\allowdisplaybreaks
\begin{eqnarray}
\label{eqGFAnsatz}
\Gamma_{5\mu}(t,q) & = & \Gamma_{5\mu}^{L}(t,q) + \Gamma_{5\mu}^{M}(t,q)\,,\\
\nonumber
\Gamma_{5\mu}^{L}(t,q) & = & \gamma_5\gamma_\mu \Sigma_A(y_+,y_-) +
2 t_\mu \gamma\cdot t \Delta_A(y_+,y_-) \\
& & + 2 i\gamma_5 \frac{q_\mu}{q^2} \Sigma_B(y_+,y_-)\,, \label{eqDSvertex}\\
\nonumber
\Gamma_{5\mu}^{M}(t,q) &=& \Delta_A(y_+,y_-)
\bigg[ \mbox{\footnotesize $\displaystyle \frac{1}{2}$} q^2 \gamma_5\gamma_\mu
+\gamma_5
[ q_\mu \gamma\cdot t
 \\
&&  -(t_\mu+ \mbox{\footnotesize $\displaystyle \frac{1}{2}$}  q_\mu )\gamma\cdot q] - i \gamma_5\gamma_\mu \sigma_{\alpha\beta} t_\alpha q_\beta\bigg] \,, \label{eqG5M}
\end{eqnarray}}
\hspace*{-0.5\parindent}where $A$, $B$ are computed in the chiral limit.  An analogous approach has proved fruitful in the study of hadron electromagnetic properties; e.g., Refs.\,\cite{Cloet:2008re,Nicmorus:2010sd,Chang:2011tx,Cloet:2013gva,Chang:2013nia,Roberts:2013mja}.

The first piece, $\Gamma_{5\mu}^{L}$ in Eq.\,\eqref{eqDSvertex}, was proposed elsewhere \cite{Delbourgo:1979me}.  It represents the most compact solution to the longitudinal chiral-limit axial-vector WGT identity, Eq.\,\eqref{eq:longsA}, in the same sense that the so-called Ball-Chiu \emph{Ansatz} for the vector vertex \cite{Ball:1980ay} solves the associated vector WGT identity.  Notably, the $1/q^2$ singularity in $\Gamma_{5\mu}^{L}$ is real: it is the most striking part of the contribution to $\Gamma_{5\mu}$ from the dynamically generated pion pole.

The second term, $\Gamma_{5\mu}^{M}$ in Eq.\,\eqref{eqG5M}, is new.  In combination with Eq.\,\eqref{eqDSvertex} it provides a solution of the coupled longitudinal and transverse WGT identities that is minimal, in the sense that it involves only those functions which appear in the dressed-quark propagator, and is also free of spurious kinematic singularities.  It is interesting that the strength of this term is modulated by $\Delta_A$, which expresses the finite-difference derivative of the vector part of the dressed-quark self energy.  As explained in closing Sec.\,5, the error associated with using Eq.\,\eqref{eqGFAnsatz} in any computation may be gauged by comparing the result obtained thereby with that produced by independently changing the signs of the last two terms in Eq.\,\eqref{eqG5M}.

\medskip

\setcounter{equation}{0}
\renewcommand{\theequation}{A.\arabic{equation}}

\noindent\textbf{Appendix A}.
Here we list the matrix-valued tensors used in Eq.\,\eqref{eqDecomposeAV} to express the transverse part of the axial-vector vertex:
\begin{subequations}
\label{eqTransverse}
{\allowdisplaybreaks
\begin{eqnarray}
T^1_{\mu} (k,p) &=& \gamma^T_\mu\,, \\  
\label{eqT2} T^2_{\mu} (k,p)  &=& i\gamma^T_\mu(\gamma\cdot q)\,, \\ 
T^3_{\mu} (k,p) &=& i\gamma^T_\mu(\gamma\cdot t)- i\,t^T_\mu\,, \\
T^4_{\mu} (k,p)  &=& \gamma^T_\mu[\gamma\cdot t,\gamma\cdot q] - 2\,t^T_\mu \gamma\cdot q \,,\\
T^5_{\mu} (k,p)  &=& i\gamma^T_\mu(\gamma\cdot q) - 2\,i\, t^T_\mu \,,  \label{eqT5}\\
\label{eqT6} T^6_{\mu} (k,p)  &=& t^T_\mu(\gamma\cdot q)\,, \\  
T^7_{\mu} (k,p)  &=& t^T_\mu(\gamma\cdot t)\,, \\
\label{eqT8} T^8_{\mu} (k,p)  &=& i\,t^T_\mu [\gamma\cdot t,\gamma\cdot q]\,,  
\end{eqnarray}}
\end{subequations}
\hspace*{-0.5\parindent}where the superscript ``$T$'' indicates that the associated four-vector is contracted with $\mathpzc{T}_{\mu\nu}=\delta_{\mu\nu} - \hat q_\mu \hat q_\nu$.

The solution for the transverse part of the axial-vector vertex involves the scalar functions $\{\tau^Y_i\}$, which we list here:
\begin{subequations}
\label{eqtauY}
{\allowdisplaybreaks
\begin{eqnarray}
\tau_1^Y(k,p)&=&-\frac{iY_3^1}{2q^2}-\frac{iY_1^2}{q^2(k^2-p^2)},\\
\tau_2^Y(k,p)&=&-\frac{(2q^2+k^2-p^2)Y_1^1}{2q^2(k^2-p^2)(k\cdot p^2 - k^2\,p^2)}\notag
\\ &&-\frac{iY_4^1}{q^2(k^2-p^2)}-\frac{Y_2^2}{q^2(k^2-p^2)},\\
\tau_3^Y(k,p)&=&\frac{Y_1^1}{(k^2-p^2)(k\cdot p^2 - k^2\,p^2)}-\frac{Y_3^2}{2q^2},\\
\tau_4^Y(k,p)&=&\frac{iY_2^1}{2(k^2-p^2)(k\cdot p^2 - k^2\,p^2)}
\notag\\ &&+\frac{i(k^2-p^2)Y_3^1}{8q^2(k\cdot
p^2 - k^2\,p^2)}\notag
\\ &&+\frac{Y_4^2}{2q^2(k^2-p^2)},\\
\tau_5^Y(k,p)&=&\frac{Y_1^1}{(k^2-p^2)(k\cdot p^2-k^2\,p^2)}, \label{eqtauY5}\\
\tau_6^Y(k,p)&=-&\frac{2iY_2^1}{(k^2-p^2)(k\cdot p^2-k^2\,p^2)}
\notag\\ &&+\frac{i(k^2-p^2)Y_3^1}{4q^2(k\cdot p^2-k^2\,p^2)}
\notag\\ &&+\frac{iY_1^2}{2q^2(k\cdot p^2-k^2\,p^2)},\\
\tau_7^Y(k,p)&=-&\frac{3iY_3^1}{2(k\cdot p^2 - k^2\,p^2)}\notag
\\&&-\frac{iY_1^2}{(k^2-p^2)(k\cdot p^2 - k^2\,p^2)},\\
\tau_8^Y(k,p)&=&-\frac{3iY_4^1}{2(k^2-p^2)(k\cdot p^2-k^2\,p^2)}
\notag\\ &&-\frac{Y_2^2}{2(k^2-p^2)(k\cdot p^2-k^2\,p^2)}
\notag\\ &&-\frac{(k^2-p^2)Y_3^2}{8q^2(k\cdot p^2-k^2\,p^2)},
\end{eqnarray}}
\end{subequations}
\hspace*{-0.5\parindent}where we have suppressed the arguments of $\{Y_i^{1,2}(k,p),i=1,\ldots,4\}$.

\medskip

\noindent\textbf{Acknowledgments}.
We are grateful to L.~Chang for providing the computer codes that enabled us to compare RL results for the axial-vector vertex with those produced by the DB kernel; and for insightful comments from I.\,C.~Clo{\"e}t and J.~Segovia.
SXQ is grateful for encouragement from D.\,H.~Rischke and acknowledges support from the Alexander von Humboldt Foundation via a Research Fellowship for Postdoctoral Researchers; and CDR acknowledges support from an \emph{International Fellow Award} from the Helmholtz Association.  This work was otherwise supported by:
University of Adelaide and Australian Research Council through grant no.~FL0992247;
Department of Energy, Office of Nuclear Physics, contract no.~DE-AC02-06CH11357;
and For\-schungs\-zentrum J\"ulich GmbH.

\medskip



\end{document}